\documentstyle[editedbook,epsfig,psfig,epsf]{mq}

\def\etal{{\em et al.} }

\def\msun{$M_{\odot}$ }

\def\cm2{cm$^2$ }
\def\se1{s$^{-1}$ }

\begin{opening}
\title{Variable gamma-ray emission from microblazars}
\author{M.M. Kaufman Bernad\'o$^1$, G.E.Romero$^{1}$ \& I.F.Mirabel$^{2,3}$}
\institute{$^1$ Instituto Argentino de Radioastronom\'{\i}a, C.C.5, (1894) Villa
Elisa, Buenos Aires, Argentina.\\
$^2$ CE Saclay, DSM/DAPNIA/Service d'Astrophysique, F91191  Gif sur Yvette
France.\\
$^3$ Instituto de Astronom\'{\i}a y F\'{\i}sica del Espacio (IAFE), C.C.67, Suc 28,
Buenos Aires, Argentina.\\
}
\end{opening}
\runningtitle{Variable gamma-ray emission from microblazars}
\runningauthor{Kaufman Bernad\'o, Romero \& Mirabel}
\begin{document}
\vspace{-0.5cm}
\begin{abstract}
{\small We propose a model for gamma-ray emitting microblazars
based on the Compton interaction of a relativistic
electron-positron plasma, ejected in a jet feature, with the
UV-photon field provided by a high-mass stellar companion. Taking
into account the gravitational effects of the star upon the
accretion disk, we predict a jet precession which results in a
variable, periodic, high-energy gamma-ray source. The specific
case of Cygnus X-1 is briefly discussed.}
\end{abstract}
\section{Introduction}
The third EGRET catalog \cite{Hartman} contains approximately 170
unidentified gamma-ray sources. A significant number of them seem
to be Population I objects \cite{Romero}. Two main groups of
sources are distinguished at mid and low galactic latitudes, one
related to the Gould belt region \cite{Grenier} and the other formed
by brighter sources at lower latitudes \cite{Gehrels}. These
sources are suspected to be galactic and their possible
counterparts include early-type stars (both isolated and in binary
systems), accreting neutron stars, radio-quiet pulsars,
interacting SNRs, and microquasars \cite{Paredes}.

We will concentrate on the last possibility, proposing a model for
high-energy emission in microquasars with jets forming a small
angle with the line of sight. These \emph{microblazars} are
expected to have highly variable and enhanced non-thermal flux due
to Doppler boosting \cite{Mirabel}, \cite{Georganopoulos 1}.

\section{Model}

Let us consider high-mass binaries with disk accretion onto the
compact object (black hole). Twin relativistic $e^+e^-$-pair jets
are ejected in opposite directions. The relativistic plasma is
injected into the external photon fields generated by the
accretion disk, the corona, and the companion star (typically UV
photons), producing inverse Compton (IC) high-energy emission. The
resulting IC specific luminosity in the case of external
monoenergetic and isotropic photon fields is given by
\cite{Georganopoulos 2}:
\begin{equation}
\frac{dL}{d\epsilon\ d\Omega} \approx D^{2+p} \frac{k V \sigma_T c
U 2^{p-1}}{\pi \epsilon_0 (1+p)(3+p)}\left(
\frac{\epsilon}{\epsilon_0}\right) ^{-(p-1)/2} \label{eq:eq1}
\end{equation}
where $D = [\Gamma (1 - \beta cos \phi) ]^{-1}$ is the Doppler
factor, $\phi$ the viewing angle, $\epsilon$ the final photon
energy, $\sigma_T$ the Thomson cross section, $U$ the energy
density of the photon field, $k$ is a constant, and $p$ is the
spectral index of the particle energy distribution given by
$N(E)\propto E^{-p}$.

\begin{figure}[htb]
\begin{center}
\includegraphics[scale=0.33]{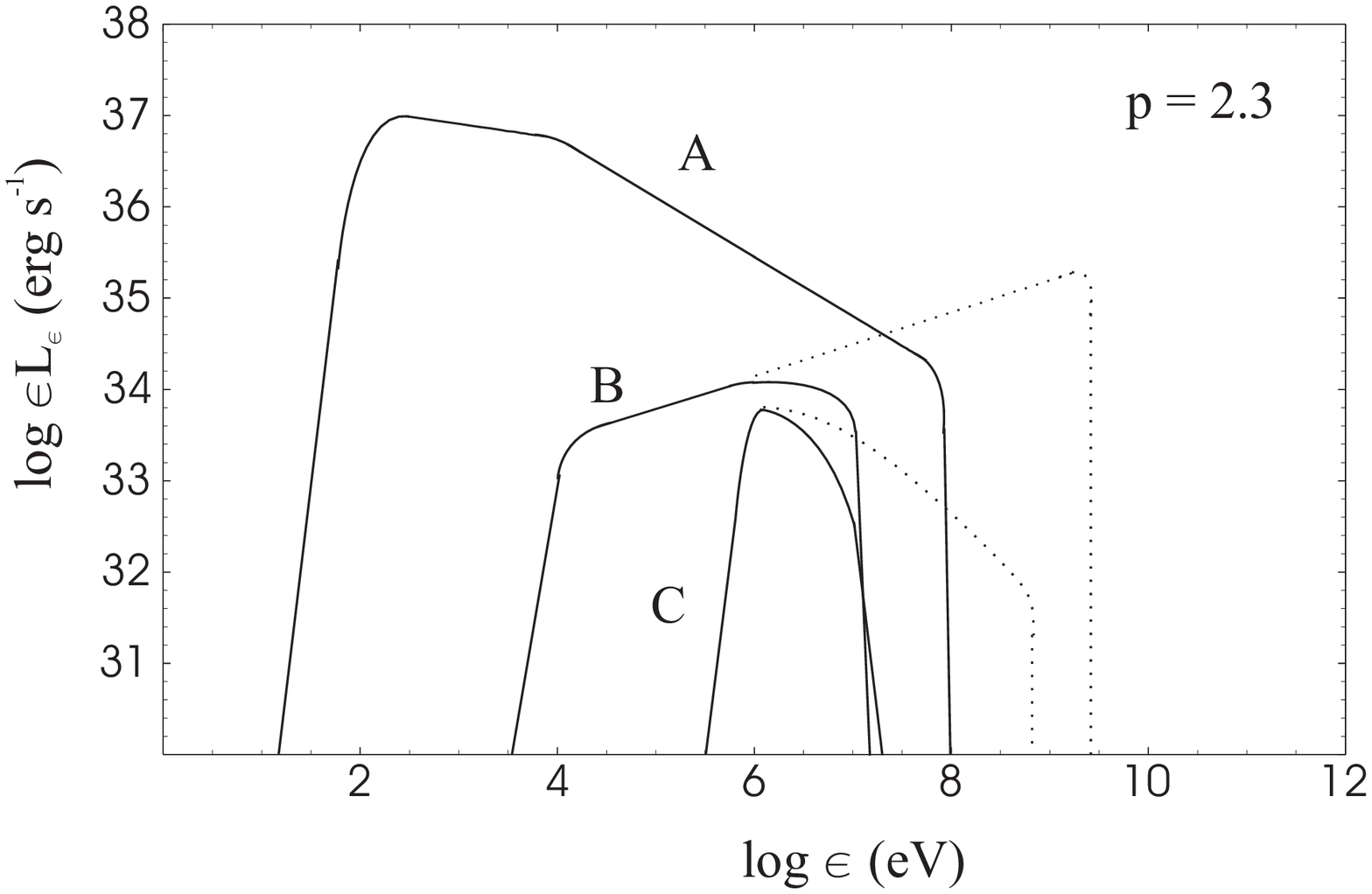} \quad
\includegraphics[scale=0.36]{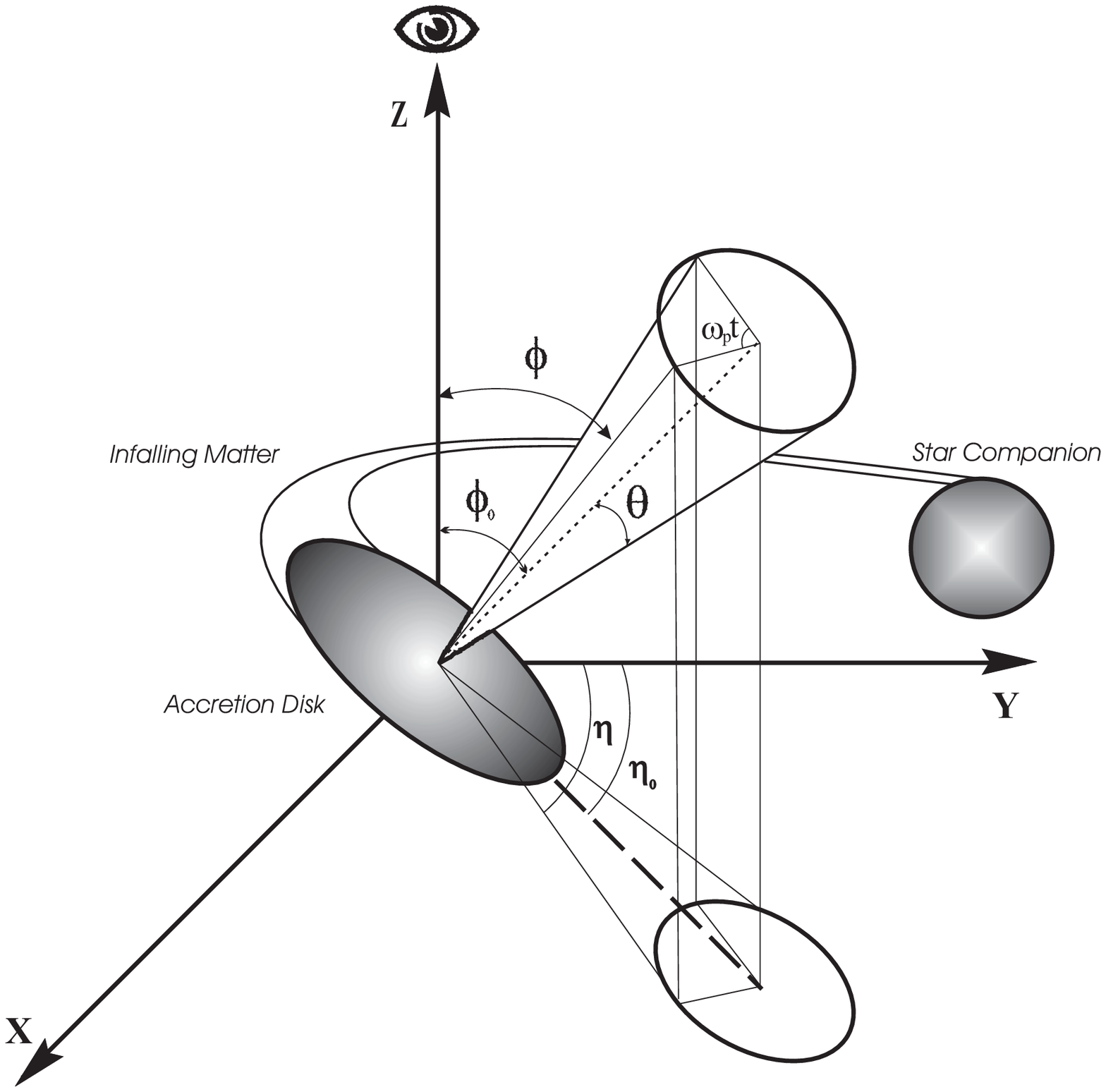} \\
\caption{\emph{Left}: Results of the model for an injection
electron spectrum with index $p=2.3$ in a cylindrical jet forming
a viewing angle of 30 degrees. The bulk Lorentz factor is
$\Gamma=5$. Three different component are shown, resulting from
the up-scattering of star (A), disk (B), and corona (C) photons.
Radiation absorbed in the local photon fields is shown in dashed
lines. Parameters for the thermal photon fields correspond to the
case of Cygnus X-1. \emph{Right}: Precessing jet model}
\label{fig:fig1} 
\end{center}
\end{figure}

\emph{Figure \ref{fig:fig1} left panel} shows the results obtained
from the calculation of the spectral energy distribution for a
specific object with $p=2.3$. Notice that eq. (\ref{eq:eq1}) has
to be modified by a factor $(1 - cos \phi_0)^{(p+1)/2}$ in the case of
interactions with disk photons, which arrive from behind the jet.
In the coronal case, the resulting spectrum is not a power-law
because of the Klein-Nishina effects \cite{Georganopoulos 2}.
Details of the calculations can be found in \cite{Romero et al.}.

The companion star in a high-mass microblazar system not only
provides a photon field for IC interactions, but also a
gravitational field that can exert a torque onto the accretion
disk around the compact object. The effect of this torque, in a
non-coplanar system, is to induce a Newtonian precession of the
disk. If the jets are coupled to the disk, as it is usually
thought \cite{Falcke}, then the
precession will be transmitted to them (a sketch of this situation
is presented in \emph{Figure \ref{fig:fig1} right panel}).

We can then introduce a time-parametrization of the jet viewing
angle $\phi(t)$ \cite{Abraham}. In this way, the boosting
amplification factor given in eq. (\ref{eq:eq1}) will periodically
oscillate as a function of time. A very weak, otherwise undetected
gamma-ray microblazar can increase its flux due to the precession
and then enter within the sensitivity of an instrument like EGRET,
producing a variable unidentified gamma-ray source (see
\cite{Kaufman et al.} for details).

\section{Cygnus X-1}

This object is a potential candidate for the proposed model. Cyg
X-1 is a black hole candidate with M $\sim$ 10 \msun and a
luminous  high mass companion (an O9.7 Iab star with L
$\sim10^{39}$ erg/sec). It presents a continuous non-thermal radio
jet \cite{Stirling} with an apparent bending that could be
attributed to precession. The source has been repeatedly detected
at MeV gamma-ray energies by the interplanetary network
\cite{Golenetskii} and by BATSE \cite{Schmidt}.

Calculations of the high-energy (kev-MeV) non-thermal spectrum,
taking into account jet interactions with photons from the stellar
companion, the accretion disk and the corona have been done (see
\emph{Figure \ref{fig:fig1}}). It can be observed that the major
contribution to the total luminosity is from the IC scattering of
stellar photons. For a viewing angle of $\sim30$ degrees and a
half-opening angle of the precession cone of $\sim16.5$ degrees,
there is a variation of about 1 order of magnitude in the emission
measured in the observer's frame because of the precession. This
means that when the jet is closer to the line of sight, the
maximum non-thermal luminosity can reach values of $\sim 10^{38}$
erg/sec, as observed \cite{Golenetskii}.

\section{Future prospects}

Most of the gamma-rays produced within the coronal region will be
absorbed by pair creation. The annihilation of these pairs would
produce a broad, blue-shifted feature in the MeV spectrum. Soon
operating INTEGRAL satellite will be able to probe Cygnus X-1
spectrum and its temporal evolution at this range of energies,
providing new tools to evaluate and constrain models like the one
presented here. AGILE and GLAST GeV observations will also help to
establish the high-energy cut-off of the injected particle
spectrum, shedding light on the energetics of the jet.

\section*{Acknowledgments}

This research was supported by Fundaci\'on Antorchas, CONICET,
ANPCT, and the ECOS Argentinian-French Cooperation Agreement.

\noindent GER thanks M. Georganopoulos for valuable discussions.

\end{document}